\documentclass[a4paper,11pt,twoside]{article}

\usepackage{jheppub} 

\usepackage{amsmath}
\usepackage{amssymb}
\usepackage{color}
\usepackage{graphicx}
\usepackage{epstopdf}  % to use eps with Typeset -> Pdftex
\usepackage[]{units} % contains \nicefrac
\usepackage{bbold} % to obtain the identity matrix as \mathbb{1} (and R, C etc)
\usepackage{amsthm} % to use theorems & C

\usepackage[numbers,sort&compress]{natbib}

\usepackage{hyperref}
\definecolor{darkblue}{cmyk}{1,0.5,0,0.2}
\hypersetup{colorlinks, bookmarksnumbered, citecolor=darkblue, linkcolor=darkblue, pdfstartview=FitH, urlcolor=darkblue, linktocpage}

\newcommand{\TeV}{\,\mathrm{TeV}}
\newcommand{\GeV}{\,\mathrm{GeV}}
\newcommand{\MeV}{\,\mathrm{MeV}}

\newcommand{\fracwithdelims}[4]{\left#1 \frac{#3}{#4} \right#2}
\newcommand{\ord}[1]{\mathcal{O}\left( #1 \right)}

\newcommand{\vev}[1]{\left\langle #1\right\rangle}
\newcommand{\Eq}[1]{Eq.~(\ref{eq:#1})}
\newcommand{\eq}[1]{eq.~(\ref{eq:#1})}

\newcommand{\eqs}[1]{eqs.~(\ref{eq:#1})}

\newcommand{\omi}[1]{{\small\ttfamily\bfseries}}

\DeclareMathOperator{\tr}{Tr}

\newcommand{\fb}{\,\mathrm{fb}}
\newcommand{\Gaa}{\Gamma_{\gamma\gamma}}
\newcommand{\Ggg}{\Gamma_{gg}}
\newcommand{\saa}{\sigma_{\gamma\gamma}}
\newcommand{\lmin}{\lambda_\text{m}}
\newcommand{\geff}{g_\text{eff}}

\newcommand{\be}{\begin{equation}}
\newcommand{\ee}{\end{equation}}
\newcommand{\bea}{\begin{eqnarray}}
\newcommand{\eea}{\end{eqnarray}}

\newlength{\myem}
\newcounter{mysubequation}[equation]

\newcommand{\SISSA}{SISSA/ISAS and INFN, I--34136 Trieste, Italy}
\newcommand{\ICTP}{ICTP, Strada Costiera 11, I--34151 Trieste, Italy}

\newcommand{\titletext}{A closer look to the sgoldstino interpretation \\ of the diphoton excess} 

\newcommand{\abstracttext}{We revisit the sgoldstino interpretation of the diphoton excess in the context of gauge mediation. While the bound on the gluino mass might seem to make the sgoldstino contribution to the diphoton excess unobservable, we show that the interpretation is viable in a thin, near critical region of the parameter space. This regime gives rise to drastic departures from the standard gauge mediation picture. While the fermion messengers lie in the (10--100)$\TeV$ range, some scalar messengers are significantly lighter and are responsible for the sgoldstino production and decay. Their effective coupling to the sgoldstino is correspondingly enhanced, and a non-perturbative regime is triggered when light and heavy messenger masses differ by a factor $\sim4\pi$. 
We also comment on the possible role of an $R$-axion and on the possibility to decouple the sfermions in this context. }

\title{\titletext}

\author[a]{P. Baratella,}
\author[a]{J.~Elias-Mir\'o,}
\author[a]{J.~Penedo,}
\author[a,b]{and A.~Romanino}

\affiliation[a]{\SISSA}
\affiliation[b]{\ICTP}

\emailAdd{pbaratel@sissa.it}
\emailAdd{jelias@sissa.it}
\emailAdd{jpenedo@sissa.it}

\begin{document}

\abstract{\abstracttext}

% \bigskip
\maketitle
%\begin{abstract}\normalsize\noindent
%\abstracttext
%\end{abstract}\normalsize\vspace{\baselineskip}

\flushbottom

% \clearpage

% \noindent

% \tableofcontents

\section{Introduction}

The excess in the diphoton channel recently reported by the ATLAS~\cite{ATLASss} and CMS~\cite{CMS:2015dxe} experiments at the Large Hadron Collider (LHC) at an invariant mass of approximately $750\GeV$ has prompted a variety of possible interpretations. The interpretation in terms of the production and decay of a sgoldstino~\cite{Petersson:2015mkr,Bellazzini:2015nxw,Demidov:2015zqn,Casas:2015blx,Ding:2016udc} places the underlying new physics into the wider context of supersymmetric extensions of the Standard Model (SM), thus going beyond the mere parameterisation of the effect in terms of ad hoc dynamics. 
%and providing a partial answer to the thorny question ``who ordered that?''. 
In fact, the very dynamics responsible for the generation of gluino and photino masses through the $F$-term of the goldstino superfield also provide, as a consequence of supersymmetry, a gluon and a photon decay width for the sgoldstinos. 

The connection between gaugino masses and decay widths is most easily illustrated in terms of an effective description of the interaction between the goldstino superfield and the SM gauge superfield strengths $W^\alpha_a$ (the index $a$ labels the different gauge fields),
\begin{equation}
 \mathcal{L}_{\text{eff}}=\frac{c_a}{\Lambda}\int d^2\theta \, X W_a^\alpha W^a_\alpha \, ,
\end{equation}
where $\Lambda$ represents the scale at which the effective operator is generated and the dimensionless coefficient $c_a$ takes everything else into account. If $X$ is the only superfield getting an $F$-term, its fermion component is the goldstino (the Goldstone of supersymmetry breaking) and its scalar partner is the sgoldstino. In terms of the $F$-term vev $F$, the gaugino masses are given by $M_a = 2 c_a F/\Lambda$, and
\begin{equation}
\mathcal{L}_{\text{eff}}= \frac{M_a}{2F}\int d^2\theta \, X W^\alpha_a W^a_\alpha = \frac{M_a}{2} \lambda_a \lambda_a + \frac{M_a}{2\sqrt{2}F} \left( s\, v^{\mu\nu}_av^a_{\mu\nu} - a\, v^{\mu\nu}_a \tilde v^a_{\mu\nu}\right)  + \ldots,
\label{eq:effective}
\end{equation}
where $\lambda_a$ and $v^{\mu\nu}_a$ are the gauginos and gauge field strengths associated to $W^\alpha_a$, and the scalar component of $X$ has been decomposed in its real and imaginary parts. 

As we show below, the effective description in \eq{effective} can hardly account for the diphoton excess in the context of concrete UV completions, in particular if gaugino masses originate from gauge mediation. The problem is not apparent in the effective description, especially if the coefficient of the effective operator is expressed in terms of $M_a/F$, as in \eq{effective}. The way out we present in this paper requires supersymmetry to be badly broken in the dynamics underlying the effective interaction, $\sqrt{F}\sim \Lambda$, in which case the relevant effects are not captured by the effective description in \eq{effective}, valid for $\sqrt{F}\ll \Lambda$. The problem, and its solution, are discussed in Sect.~\ref{sec:origin} (after a few preliminaries in Sect.~\ref{sec:naiveEFT}), in the simple case in which the operator in \eq{effective} originates from a loop of chiral messengers, as in minimal gauge mediation. The interpretation of the excess relies on the onset of a non-perturbative regime, when approaching a critical point where one of the scalar messengers becomes light. In Sect.~\ref{sec:speculations} we speculate on the possible role of such a regime in simple models of supersymmetry breaking. In Sect.~\ref{sec:collimated} and~\ref{sec:D} we comment on the possible role of an $R$-axion and of $D$-terms raising the sfermion masses well above the experimental limits. In Section~\ref{sec:conclusions} we summarize and conclude.

\section{The effective description}
\label{sec:naiveEFT}

In this section we take  \eq{effective} at face value and  show how the size of the diphoton excess translates into constraints on its parameters. 
A similar derivation was done in Refs.~\cite{Petersson:2015mkr,Bellazzini:2015nxw,Demidov:2015zqn,Casas:2015blx,Ding:2016udc}.

Although, needless to say, the very existence of the anomaly is not yet established, we will assume that it corresponds to the production of a scalar resonance, identified with a scalar component of $X$, decaying into two photons. The interaction in \eq{effective} provides the necessary ingredients for the production of the resonance through gluon fusion and its decay into photons.  We  consider a reference value of  $\saa\equiv\sigma(pp\to s \to \gamma\gamma) =8\fb$ for the cross section at $13\TeV$, see e.g.~\cite{Buttazzo:2015txu,Knapen:2015dap,Franceschini:2015kwy,Gupta:2015zzs,Falkowski:2015swt,Buckley:2016mbr}.   In the light of the presently uncertain experimental situation, we do not aim at accounting for a possibly large width of the resonance. 

As we will see, obtaining a large enough partial width $\Gamma(s\to\gamma\gamma) \equiv \Gaa$ is not at all trivial. Let us then conservatively consider the minimum value of $\Gaa$ necessary to account for the anomaly. It is easy to see that such a minimum value is obtained when i) $\gamma\gamma$ and the partons $pp$ involved in the production are the only decay channels, so that $\Gamma_\text{tot} = \Gaa+\Gamma_{pp}$, ii) $\Gamma_{pp}$ dominates the width, and iii) the resonance is produced through gluon fusion ($pp = gg$). Which happen to be quite plausible conditions. Then, one gets~\footnote{We have used a $K$ factor $K_{gg}\approx2.8$, as in~\cite{Aparicio:2016iwr}; the Higgs cross-section $\sigma_{13\text{TeV}}(gg\rightarrow H(750~\text{GeV}))\approx736~\text{fb}$~\cite{higgsxswg}; and $C_{gg}=2137$~\cite{Franceschini:2015kwy} for the   gluon parton distribution function from NLO MSTW 2008~\cite{Martin:2009iq}.} 
\begin{equation}
\Gamma(s\to \gamma\gamma) \approx 0.4\MeV \fracwithdelims{(}{)}{\saa}{8\fb} .
\label{eq:Gmin}
\end{equation}
In terms of the effective interaction in \eq{effective}, the prediction for the photon partial width is~\cite{Perazzi:2000ty}
\begin{equation}
\Gamma(s\to \gamma\gamma) = \frac{m^3_s M^2_\gamma}{32\pi F^2} , 
\qquad
M_\gamma = c^2_W M_1 + s^2_W M_2,
\label{eq:Gaaeff}
\end{equation}
where $m_s \approx 750\GeV$ is the mass of the resonance and  $M_\gamma$ is expressed in terms of the bino and wino masses $M_1$ and $M_2$,  and the Weinberg angle $\theta_W$. Comparing \eqs{Gmin} and~(\ref{eq:Gaaeff}) we obtain 
\begin{equation}
\sqrt{F} \lesssim 4\TeV \fracwithdelims{(}{)}{M_\gamma}{200\GeV}^{1/2} \fracwithdelims{(}{)}{8\fb}{\saa}^{1/4} .
\label{eq:Fmax}
\end{equation}
In this effective approach, the size of the diphoton excess points to a very low scale of supersymmetry breaking. It is not easy to deal with such a low scale, as we expect gauge mediation to provide the main source of gaugino masses at this scale and gaugino masses to be loop suppressed, as we discuss in the next Section. 

\section{A simple ultraviolet completion}
\label{sec:origin}

Let us now discuss in greater detail the interpretation of the diphoton anomaly taking into account the origin of the effective interaction in \eq{effective}. We assume that gaugino masses are  obtained at the one loop level through the exchange of messenger superfields, directly coupled to supersymmetry breaking, in vectorlike representation of the SM group $G_\text{SM}$, as in minimal gauge mediation. Note that on general grounds~\cite{Angelescu:2015uiz,Backovic:2015fnp,Ellis:2015oso,Bellazzini:2015nxw,Gupta:2015zzs,Low:2015qep,Dutta:2015wqh,Kobakhidze:2015ldh,No:2015bsn,Chao:2015ttq,Fichet:2015vvy,Curtin:2015jcv,Falkowski:2015swt}, the interpretations of the diphoton anomaly also requires the existence of vectorlike fields, on top of the $750\GeV$ resonance, mediating its production and decay. The gauge mediation messengers play precisely that role, thus providing a wider context for the existence of the vectorlike fields as well. 
% In this Section, we discuss the interpretation of the diphoton anomaly in this setup. 

To be specific, we add to the minimal supersymmetric SM (MSSM) field content a chiral superfield $X$, with non-vanishing scalar and $F$-term vevs
\begin{equation}
X = x + \sqrt{2} \psi \theta + f \theta^2, 
\qquad \vev{X} = M + F\theta^2,
\qquad x = M + \frac{s+ia}{\sqrt{2}}. 
\label{eq:X}
\end{equation}
The vev of $X$ plays the role of the supersymetry breaking spurion of minimal gauge mediation ($M$ and $F$ can be taken positive without loss of generality). On top of that, the dynamical degrees of freedom of $X$ also play a role here. In particular, the $750\GeV$ resonance will be associated to the real scalar $s$. Both $s$ and $a$ are assumed for simplicity to be mass eigenstates. The origin of the supersymmetry breaking masses of $s$ and $a$ is a model-depedent issue, which we will not investigate in this Section. As mentioned, if $F$ were the only source of supersymmetry breaking, $\psi$ would be the goldstino and $x$ the sgoldstino. We also add messenger superfields $\Phi_i$, $\bar{\Phi}_i$ in irreducible, conjugated (possibly real) representations of the SM group.  In order to generate one-loop masses for the three gauginos, the messengers should have non-trivial transformations under all the three SM gauge factors. 
 They are coupled to supersymmetry breaking through $X$ only,
\begin{equation}
\mathcal{L}_\Phi =\int d\theta^2\, \lambda_i X \Phi_i \overline{\Phi}_i + \text{h.c.},
\label{eq:W}
\end{equation}
where the coupling can be taken diagonal and positive without loss of generality. In the following, we will denote by $\lmin$ the minimum value of the couplings $\lambda_i$ (at the messenger scale).

Gaugino masses arise from \eq{W} through the standard gauge mediation mechanism. Sfermion masses also get a contribution from \eq{W}, which however is not necessarily the only, nor the dominant, one (see e.g.\ Sect.~\ref{sec:D}). We will therefore assume that the sfermions do not play a role in the diphoton anomaly. 

Let us now discuss whether the diphoton excess can be accounted for in this setup. Before discussing the scenario we are interested in, in which $F\sim \lambda M^2$, we show that this is not possible in the $F\ll \lambda M^2$ limit. 

% In the rest of the section we discuss two possible hierarchies, $F\ll \lmin M^2$ and $F\sim \lmin M^2$.

\subsection{$F \ll \lambda M^2$}
\label{sec:Fsmall}

In the  limit where $F \ll \lmin M^2$, supersymmetry breaking can be neglected when integrating out the messengers $\Phi_i+\overline{\Phi}_i$, whose fermion and scalar components all have masses close to $\lambda_i M$. The effective interaction in \eq{effective} follows, with 
\begin{equation}
\frac{c_a}{\Lambda} = \frac{\alpha_a}{8\pi M} N_a \, ,
\quad\quad \text{giving} \quad\quad
M_a = \frac{\alpha_a}{4\pi} \frac{F}{M} \, N_a \, ,
\label{eq:matching}
\end{equation}
where  $a$ labels the factor of the SM group (U(1)$_Y$, SU(2)$_L$, SU(3)$_c$ respectively), and $N_a = \sum_i N_{a,i}$, and $N_{a,i}$ is the corresponding Dynkin index of $\Phi_i + \overline{\Phi}_i$. For instance, if the messengers form complete SU(5) multiplets, $N_a=1$ for a $\mathbf{5}+\overline{\mathbf{5}}$ and $N_a=3$ for a $\mathbf{10}+\overline{\mathbf{10}}$. 

It is now easy to show that the numerical results obtained in the previous Section are not phenomenologically viable in this context. The explicit expression of the photino mass is 
\begin{equation}
M_\gamma = \frac{\alpha}{4\pi}\frac{F}{M} \, N_\gamma,
\qquad
N_\gamma = 2\tr(Q^2) ,
\label{eq:photino}
\end{equation}
where $\alpha = e^2/(4\pi)$ and the Dynkin $N_\gamma$ is obtained tracing on the $\Phi_i$ superfields only. Plugging the above expression in \eq{Gaaeff}, the dependence on $F$ drops out, 
\begin{equation}
\Gamma(s\to \gamma\gamma) = \frac{m^3_s}{M^2} \frac{\alpha^2}{(8\pi)^3}N^2_\gamma \, . 
\label{eq:Gaaeffmess}
\end{equation}
\Eq{Gmin} then gives an upper limit on the messenger scale
\begin{equation}
M \lesssim 61\GeV \, N_\gamma \fracwithdelims{(}{)}{8\fb}{\saa}^{1/2} 
\quad \Rightarrow\quad
\lmin N_\gamma \gtrsim 16 \, \frac{M_{m}}{\text{TeV}} \fracwithdelims{(}{)}{\saa}{8\fb}^{1/2} , 
\label{eq:Mmax}
\end{equation}
where $M_{m} = \lmin M$ is the mass of the lightest messengers. The experimental bounds on the latter\footnote{The experimental bounds on the mass of the lightest messengers depend on their decay mode. One possibility is that they decay through a small coupling to the MSSM fields. For example, this is the case if the presence of a superpotential interaction $W_\text{LQ} = \bar\Phi Q L$. Then, at TeV energies these interactions lead to lepto\-quark type couplings involving the lightest fields $\phi_l$. The lower bounds on the their mass are around $650~\GeV$~\cite{Aad:2015caa,Khachatryan:2015bsa}. In various instances below we consider higher multiplicities in the number of messengers, therefore leading to higher multipliciites  of the leptoquark couplings. 
%In the future it would be interesting to perform a dedicated collider study on such type of searches involving the messenger fields, at the moment though, we find it reasonable to assume bounds up to $\geq 1-2$~TeV in the various instances studied below.
We will therefore use $1\TeV$ as a reference lower bound.}  
%(on the order of $0.5-1$~TeV at least, see next section) 
require relatively large values of the Dynkin index. Such values can be achieved in the presence of a rich enough set of messengers. For example, a full family of messengers, filling a $\overline{\mathbf{5}}+\mathbf{10}$ representation of SU(5), would give $N_\gamma = 32/3$,  and $\lmin > 1.5$ would suffice to allow TeV scale messengers. 

The problem arises from the gaugino masses, the gluino mass $M_3$ in particular. Being loop suppressed with respect to the messenger masses $M_m$, the experimental bound on $M_3$ forces $M_m$ to be in the $\ord{100\TeV}$ region, barring unrealistic values of $N_3$. 
%In fact, using that $F\ll \lambda_i M^2 = M M_{m, i }$, we have
We have in fact
\begin{equation}
M_3 = \frac{\alpha_3}{4\pi}\frac{F}{M} \, N_3 \ll \frac{\alpha_3}{4\pi} \, M_m \, N_3 
%\lesssim 0.5\GeV (\lambda N_3 N_\gamma) \fracwithdelims{(}{)}{8\fb}{\saa}^{1/2} ,
\quad \Rightarrow \quad
M_{m} \gg \frac{130 M_3}{N_3} .
\label{eq:M3maxeff}
\end{equation}
When plugged in \eq{Mmax}, such large messenger masses require unrealistic values of $N_\gamma$. 

In the expressions above, we have assumed that only $s$ contributes to the diphoton anomaly. The possibility that both $s$ and $a$ contribute is often considered, also in connection to the possibility of explaining a possibly sizeable width of the $750\GeV$ resonance~\cite{Petersson:2015mkr,Casas:2015blx,Ding:2016udc}. The presence of both contributions would enhance the photon width by a factor of two, but would not change our conclusions. 

\subsection{$F \sim \lambda M^2$}

Drastic departures from the grim predictions of the previous subsection arise in the regime in which supersymmetry breaking is sizeable, and the effective description in \eq{effective} does not apply. In order to obtain the expressions for the partial widths $\Gaa$, $\Ggg$ in this regime, we first write the relevant interactions. Omitting for simplicity the messenger index $i$, the mass terms for the fermion ($\psi$, $\bar{\psi}$) and scalar ($\phi$, $ \bar{\phi}$) components of the messengers $\Phi$, $\bar{\Phi}$ are
\begin{equation}
-\mathcal{L}_\text{mess}^{(2)} = 
\left(
\lambda M \psi \bar \psi + \text{h.c.}
\right) +
\lambda
(\phi^\dagger, \bar\phi )
\begin{pmatrix}
\lambda M^2 & F \\
F & \lambda M^2
\end{pmatrix}
\begin{pmatrix}
\phi \\
\bar{\phi}^\dagger
\end{pmatrix} .
\label{eq:mmass}
\end{equation}
The fermion messengers $\psi$ and $\bar\psi$ form Dirac spinors with mass $M_m = \lambda M$.  The scalar mass eigenstates are 
\begin{equation}
\phi_{h,l} = \frac{\phi \pm \bar{\phi}^\dagger}{\sqrt{2}}, 
\quad
\text{with masses}
\quad
m^2_{h,l} = M^2_m \pm \lambda F. 
\label{eq:hl}
\end{equation}
We assume to be in the regime $F \leq \lmin M^2$, so that no messenger develops a vev.

The partial widths $\Gaa$, $\Ggg$ and the gaugino masses $M_a$ are generated by the gauge interactions and the trilinear messenger interactions
\begin{equation}
-\mathcal{L}_\text{mess}^{(3)} = 
\left(
\lambda \frac{s+i a}{\sqrt{2}} \psi \bar \psi + \text{h.c.}
\right) + 
\sqrt{2}\lambda^2 M s \,
(|\phi_l|^2 + |\phi_h|^2) .
\label{eq:trilinear}
\end{equation}
In particular, the decay widths of $s$ (but not of $a$) get a contribution from trilinear interactions with the scalar messengers with masses $m^2_{h,l} = \lambda^2 M^2 \pm \lambda F$. Parametrically, the strength of the trilinear coupling of the lightest messenger is measured by the effective coupling
\begin{equation}
\geff =  \lambda\, \frac{M_m}{m_{l}} = 
\frac{\lambda^2 M}{(\lambda^2 M^2 - \lambda F)^{1/2}} .
\label{eq:geff}
\end{equation}
The crucial observation is that there exists a small region of the parameter space, near the critical point $M = F/\lambda$, where the lighter scalar messenger is significantly lighter than its natural scale $M_m$, and its effective coupling to the scalar resonance $s$ is correspondingly enhanced. This is the regime in which the interpretation of the diphoton excess has a chance to be phenomenologically viable, and that we will study in detail. We will call it the ``near-critical'' regime. When the enhancement becomes very large, the system enters a strongly interacting regime.  

A few comments are in order. 
\begin{itemize}
\item In the near-critical regime, supersymmetry is maximally broken, $F \approx \lambda M$, and the dynamics is far from being described, even qualitatively, by the effective approach in \eq{effective}. 
\item The near-critical region is fine-tuned, as $F$ and $\lambda M$ need to be close, with the fine-tuning parameter given by $\Delta = (M_m/m_l)^2 = (\geff/\lambda)^2$. In Sect.~\ref{sec:speculations} we will speculate on a possible dynamical origin of such a degeneracy. 
\item A lower bound to the potential is guaranteed by supersymmetry, independently of the size of the trilinear coupling. Assuming the potential is stabilised in a nearly critical point, we expect the minimum to be meta-stable. 
\item In the near-critical regime, the gaugino masses are not drastically enhanced, for given $M$ (a moderate enhancement comes from the $F\sim \lambda M^2$ corrections to the standard $F \ll \lambda M^2$ expressions). On the other hand, the diphoton anomaly is controlled by the lightest scalar messenger, and it is now possible to keep its mass light (to get a sizeable diphoton signal) while allowing $M$ to be much larger (to get a gluino mass above experimental bound). 
\item In the presence of multiple messengers, if none of the couplings $\lambda_i$ are (approximately) degenerate, only one messenger, the one with $\lambda_i = \lmin$, can benefit from an enhanced coupling. On the other hand, in the presence of an (approximate) degeneracy of different $\lambda_i$, e.g.\ consequence of a symmetry, more scalar messengers can be light at the same time. Such a degeneracy should involve messenger with same quantum numbers under the SM gauge group, as gauge radiative corrections could otherwise spoil the degeneracy. For example, in the context of unified theories, the different SM components of a unified multiplet would have equal couplings $\lambda_i$ at the grand unification (GUT) scale, but the different RGE running would lift the degeneracy at low energy. 
\end{itemize}

The broad picture that emerges has therefore: a resonance at $750\GeV$ associated to the sgoldstino $s$; a messenger scale $M_m$ of a few tens of TeV, so that the loop suppressed gluino mass can be above the experimental bound; a number of messengers (with same SM quantum numbers) with a near critical coupling $\lambda \approx F/M^2$ with an anomalously light scalar component (at the 1--2 TeV scale), responsible of the production and decay of the $\sim 750\GeV$ resonance, and with all the other components at the scale $M_m$; TeV scale gaugino masses generated by both light and heavy messengers. With this broad picture in mind, let us now proceed to a more detailed discussion. 

%In the near-critical regime, supersymmetry is maximally broken, $F \sim \lambda M$, and the dynamics is far from being described, even qualitatively, by the effective approach in \eq{effective}. 
%The broad picture that emerges has therefore: a resonance at $750\GeV$ associated to the sgoldstino $s$; a messenger scale $M_m$ of a few tens of TeV, so that the loop suppressed gluino mass can be above the experimental bound; a set of messengers (with same SM quantum numbers) with a near critical coupling $\lambda \approx F/M^2$ giving rise to an anomalously light scalar component (at the $1$-$2$~TeV scale)  responsible for  the production and decay of the $\sim 750\GeV$ resonance, and with all the other components at the scale $M_m$;  and  TeV scale gaugino masses generated by both light and heavy messengers. With this broad picture in mind, let us now proceed to a more detailed discussion. 

The general one-loop expressions for the partial decay widths into gluons and photons of $s$ and $a$, due to the loop of scalar and fermion messengers are given in the Appendix. There, we also provide the expressions for the decays into $ZZ$, $Z\gamma$, $WW$. The decay into two Higgses is absent at the one-loop level. In the limit in which the sgoldstino is  lighter than the messengers, $m^2_s \ll 4 M^2_m, 4 m^2_l$, those expressions become
\bea
\Gamma(s\to gg) &=& \frac{m^3_s}{M^2} \frac{4}{9} \frac{8 \alpha^2_3}{(8\pi)^3} \left|   \sum_i N_{3,i}  \left[1+ \frac{1}{4}\left(\frac{\lambda^2_i M^2}{m^2_l} + \frac{\lambda^2_i M^2}{m^2_h} \right)\right]\right|^2 \, ,  \label{eq:partial1}
\\[.2cm]
\Gamma(a\to gg) &=& \frac{m^3_a}{M^2} \frac{8 \alpha^2_3}{(8\pi)^3} \, N^2_3 \, , \label{eq:partial2} \\[.2cm]
\Gamma(s\to \gamma\gamma) &=& \frac{m^3_s}{M^2} \frac{4}{9} \frac{\alpha^2}{(8\pi)^3}\left| \sum_i N_{\gamma,i}   \left[ 1+ \frac{1}{4}\left( \frac{\lambda^2_i M^2}{m^2_l} + \frac{\lambda^2_i M^2}{m^2_h} \right)\right] \right|^2  \, ,  \label{eq:partial3} \\[.2cm]
\Gamma(a\to \gamma\gamma) &=& \frac{m^3_a}{M^2} \frac{\alpha^2}{(8\pi)^3} \, N^2_\gamma \, .
\label{eq:partial4}
\eea
As mentioned, we expect $M$ to be of order of a few tens of TeV or more, in order for the loop suppressed gluino mass to be above the experimental bounds. Eqs.~(\ref{eq:partial1})--(\ref{eq:partial4}) then show that the field $a$ cannot play a role in the diphoton anomaly, as the corresponding widths are suppressed by $M^2$. Hence in this context a possibly sizeable width of the diphoton resonance cannot be explained in terms of the production of two resonances close in mass associated to the fields $s$ and $a$~\cite{Petersson:2015mkr,Casas:2015blx,Ding:2016udc}. On the other hand, when $m_l$ is around the TeV scale, the corresponding terms in eqs.~(\ref{eq:partial1}) and (\ref{eq:partial3}) dominate and we have
\bea
\Gamma(s\to gg) &\approx& \frac{m^3_s}{m^2_l} \, \frac{4}{9} \frac{8 \alpha^2_3}{(8\pi)^3} \left( \bar N_{3} \, \frac{\lmin^2 M}{4m_l} \right)^2 \, , \label{partial00}
\\[.2cm]
\Gamma(s\to \gamma\gamma) &\approx& \frac{m^3_s}{m^2_l}  \, \frac{4}{9} \frac{\alpha^2}{(8\pi)^3} \left( \bar N_{\gamma} \, \frac{\lmin^2 M}{4m_l}
\right)^2 \, ,
\label{partial01}
\eea
where $\bar N_{3}$, $\bar N_{\gamma}$ are the Dynkin indices summed only over the messengers in the near-critical regime, i.e.\ with a light scalar degree of freedom (assumed for simplicity to have all the same mass $m_l$), which all necessarily have $\lambda_i \approx \lmin$. 

In the expressions for the partial widths in eqs.~(\ref{partial00}) and (\ref{partial01}), the (multi-TeV)$^2$ suppression $M^2$ has been replaced by  $m^2_l$, which is allowed to be close to its $\ord{\text{TeV}^2}$ experimental bounds. On top of that, the light scalar messengers further enhance the signal through the additional factor $(\lmin^2 M/m_l)^2$, which corresponds to  an  enhanced  effective coupling  $\geff$, see \eq{geff}. To get a feeling of the size of the effect, we observe that the same enhancement could be obtained in the case of a standard fermion loop (with same SM quantum numbers, mass, and in the same limit used in eqs.~(\ref{partial00}) and (\ref{partial01})) using a Yukawa coupling $\lambda_f = \lmin^2 M/(4 m_l)$ (with the Yukawa normalised as in \eq{trilinear}). 

Note that $\lmin^2 M$ cannot be taken arbitrarily larger than $m_l$, as the light messenger interactions become non-perturbative for large $\geff$. The qualitative naive dimensional analysis (NDA) estimate of the onset of the strongly interacting regime is
\begin{equation}
\geff \sim \geff^* \equiv 4\pi. 
\label{eq:gstar}
\end{equation}
As a consequence, we are only allowed to consider values of the heavy messenger scale bound by $M_m \lesssim (\geff^*/\lmin) m_l $.
Notice in particular that lowering $\lmin$ allows to rise the upper bound on $M_m$.~\footnote{We can refine the perturbativity bound of the EFT containing the sgoldstino and the $\phi_l$ field. With  the normalization $\delta{\cal L}=\sqrt{2} \lambda^2 M s |\phi_l |^2$  one gets  $\geff^*\sim 4\pi /(4dn)^{1/4}$, where $n$ is the number of messenger fields and $d$ the length of the representation. Although the bound is obtained under the assumption $m_l\gg m_s$,  a similar bound can be obtained for $m_l\ll m_s$.}

%%%%%%%%%%%

\begin{figure}[t]
\begin{center}
\includegraphics[scale=0.505]{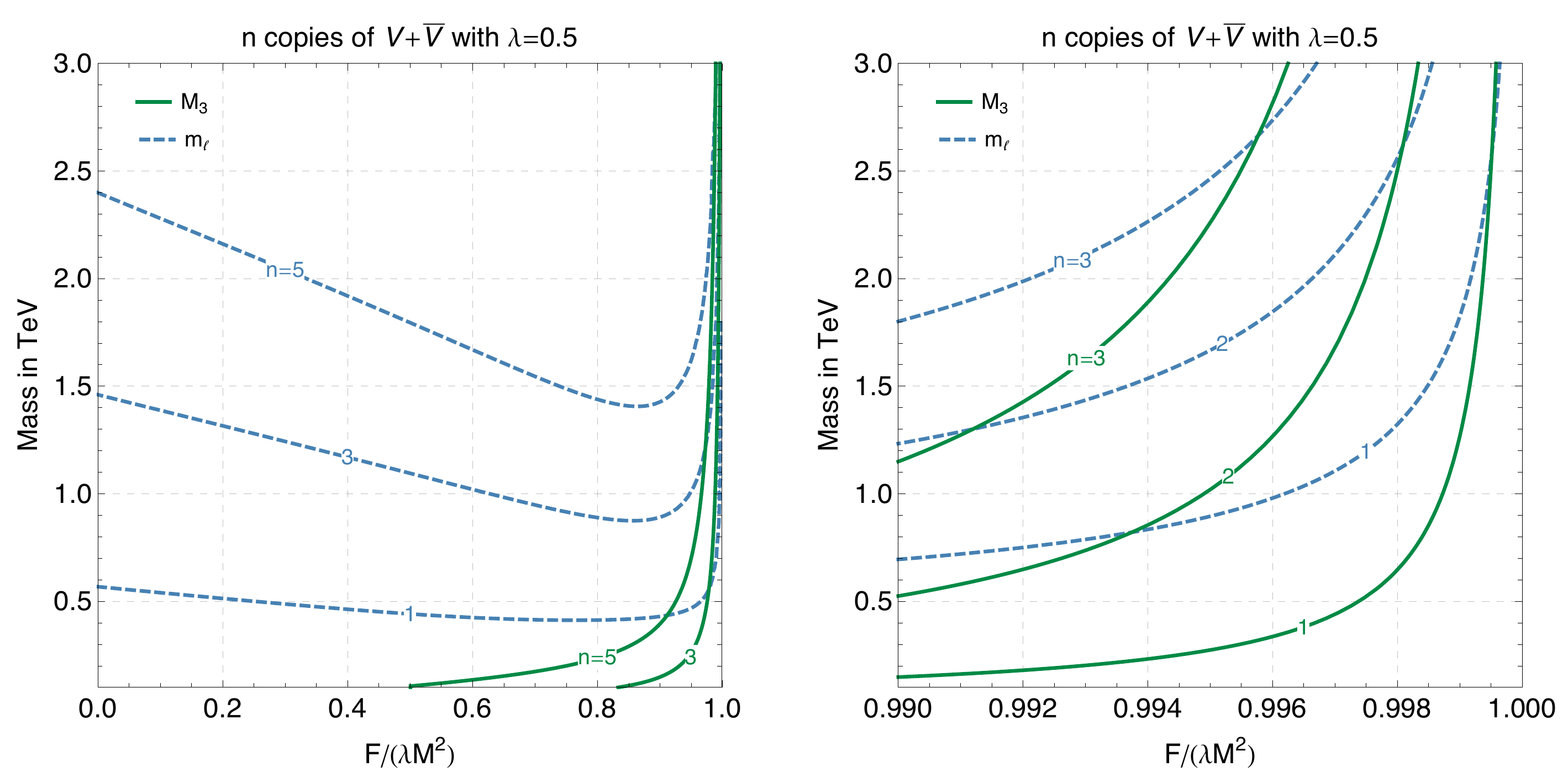}  \end{center}
\caption{\emph{ \textbf{Left:} Values of the lightest messenger mass $m_l$ (dashed blue) and gluino mass $M_3$ (green) fitting  the signal strength (with $\sigma_{\gamma\gamma} = 8\fb$), as a function of the parameter $F/(\lambda M^2)$. The $n$  messengers are assumed to be in the $  ({\bf 3},{\bf 2})_{\pm 5/6}$ SM irreps. \textbf{Right:} left plot zoomed in.} }
\label{M3vsMm}
\end{figure}

\medskip 

For completeness, let us   exemplify the discussion above in a simple setup. We consider a set of $n$ messenger pairs $V+\bar V$ with SM quantum numbers $({\bf 3},{\bf 2})_{-5/6} + ({\bf \bar 3},{\bf 2})_{5/6}$.
In the left hand side of  Fig.~\ref{M3vsMm} we show the values of the gluino mass (green) and of the mass of the  lightest set of scalar messengers $\phi_l$ (dashed blue) fitting the signal strength as a function of $F/(\lambda M^2)$. Each contour line  corresponds  to a different number of messengers, as indicated on top of each curve.  The plot is done using the full one-loop decay widths given in the Appendix. 

The lower bound on the gluino mass ($M_3\gtrsim 1.7$~TeV) requires increasing $F/M$, while $M$ needs to be increased accordingly in order to  keep a constant signal strength, see eqs.~(\ref{partial00}) and (\ref{partial01}). In fact, the plot shows that  the parameters of the theory are pushed into the near-critical regime $F/(\lambda M^2)\approx 1$ for  realistic gluino and messenger masses. In the right hand side of  Fig.~\ref{M3vsMm} we zoom into the critical region. We find that, as long as we accept a significant tuning, this toy example is able to both fit the signal and exceed the present lower bounds on scalar messengers and gluinos. 

On the other hand, the plot shows that in the effective regime, i.e.\ for $F\ll \lambda M^2$, the gluino mass cannot be accounted for. For high enough $n$, the plot in Fig.~\ref{M3vsMm} is only indicative, as the theory presents  close Landau poles in both the running of $\lambda$ and the gauge couplings, which is discussed in detail in  Sec.~\ref{sec:examples}. Also notice that for $F/(\lambda M^2)$ close enough to $1$ the trilinear coupling become non-perturbative $\geff \sim \geff^*$.

\subsection{Quantitative analysis}
\label{sec:quantitative}

Let us now show quantitatively that the diphoton excess can indeed be reproduced while keeping $m_l$ and $M_3$ above the experimental limits. As mentioned, the excess is controlled by $m_l$, while gaugino masses depend on $M_m$. We want to keep $m_l$ around the TeV scale, above its experimental bound, which helps fitting the excess. Unlike in the case of Sect.~\ref{sec:Fsmall}, this does not make the gluino mass unacceptably small, as we can now take $M_m$ significantly heavier. Using eq.~(\ref{partial01}) and~(\ref{eq:Gmin}), we find that fitting the diphoton excess requires
\begin{equation}
\frac{\geff}{\geff^*} \approx \frac{8.0}{\bar N_{\gamma}} \fracwithdelims{(}{)}{m_l}{\text{TeV}} \fracwithdelims{(}{)}{\sigma_{\gamma\gamma}}{8\fb}^{1/2} . 
\label{eq:NgammaNC}
\end{equation}
In order to keep $\geff$ below $\geff^*$, while keeping $m_l\gtrsim 1 \TeV$, the messengers with light scalar components should have $\bar N_\gamma \gtrsim 8$. We will discuss examples in Sect.~\ref{sec:examples}. Making the effective coupling $\geff$ large only gives a moderate gain with respect to the case in Sect.~\ref{sec:Fsmall}. Infact, the results for the diphoton cross section are the same (for equal mass and SM quantum numbers of the relevant degrees of freedom) when $\geff = 6\, \lmin$, and the factor 6 limits the gain. 

On the other hand, the previously hopeless situation with the gluino mass is now completely different. To start with, the standard expression in \eq{matching} for the gluino mass gets a $\ord{1}$ enhancement in the $F\sim \lambda M^2$ regime. In the near-critical region, $F\approx \lambda M^2$, the enhancement is given by a factor $\log4 \approx 1.4$. Also, extra messengers not in the near-critical regime can contribute to the gluino mass, while being negligible in the diphoton signal. We therefore have
\begin{equation}
M_3 = \frac{\alpha_3}{4\pi} M_m \bar N_3 \log 4 + \Delta M_3,
\label{eq:M3NC}
\end{equation}
where $\Delta M_3$ is the contribution of the non-critical messengers, giving
\begin{equation}
M_m \approx \frac{100\TeV}{\bar N_3} \fracwithdelims{(}{)}{M_3 - \Delta M_3}{\text{TeV}} . 
\label{eq:MmaxNC}
\end{equation}
Most important, such a large value of $M_m$ is now allowed, as long as $\geff$ it is not too large. The value of $\geff$ required by \eq{MmaxNC} is
\begin{equation}
\frac{\geff}{\geff^*} \approx \frac{8\lmin}{\bar N_{3}} \fracwithdelims{(}{)}{M_3- \Delta M_3}{m_l} .
\label{eq:N3NC}
\end{equation}
We therefore conclude that we can make $M_3$ large enough, while not exceeding the NDA bound on $\geff$, if $\bar N_3\gtrsim 8\lmin$. We will discuss examples in the next subsection. Note that smaller values of $\lmin$ help with perturbativity.

\subsection{Examples}
\label{sec:examples}

Let us discuss a few examples of viable messenger field content, leading to $\geff \lesssim \geff^*$ in eqs.~(\ref{eq:NgammaNC},\ref{eq:N3NC}). As mentioned, we prefer the near-critical messengers $\Phi_i$ to be given by $n$ copies of the same SM irreducible representation, to guarantee that the near-equality of their coupling to $X$, possibly consequence of a symmetry, is not spoiled by gauge radiative corrections. In order to induce the diphoton signal, we need them to be colored and electrically charged. Other, non-critical, messengers can also be around. Their contribution to the diphoton signal will be negligible, but they can play a role in gauge coupling unification. On the other hand, if a Landau pole for the gauge coupling is to be avoided below the GUT scale, the total number of messengers cannot be too large. 

Different model building avenues are available, depending on whether or not one aims at the perturbativity of gauge couplings up to the GUT scale and at gauge coupling unification. If the perturbativity of gauge couplings is not an issue, the bounds in eqs.~(\ref{eq:NgammaNC},\ref{eq:N3NC}) can be easily satisfied while maintaining the light scalar interactions semi-perturbative. Let us then aim at models with perturbative gauge couplings up to the GUT scale (see~\cite{Hall:2015xds, Barbieri:2016cnt} for a related discussion). This requires $N_1, N_2, N_3 \lesssim 5$, where $N_a$ are the Dynkin indices of all messengers (near-critical and not). 

If gauge coupling unification is not an issue, extra, non near-critical superfields are not required, and we can assume $\bar N_i = N_i$. Addressing the gluino mass constraints while keeping $\geff$ under control is not an issue: the UV perturbativity condition $N_3 \lesssim 5$ is always compatible with $\geff < \geff^*$, in \eq{N3NC}, for an appropriate value of $\lmin$. The only drawback of a smaller $\lmin$ is the higher fine-tuning necessary for near-criticality. As for fitting the signal strength, the relation $\bar N_\gamma = \bar N_2 + (5/3) \bar N_1$ shows that one can obtain relatively large values of $N_\gamma$ in \eq{NgammaNC} while keeping $N_1, N_2 < 5$.

\begin{figure}[t]
\begin{center}
  \includegraphics[scale=0.5205]{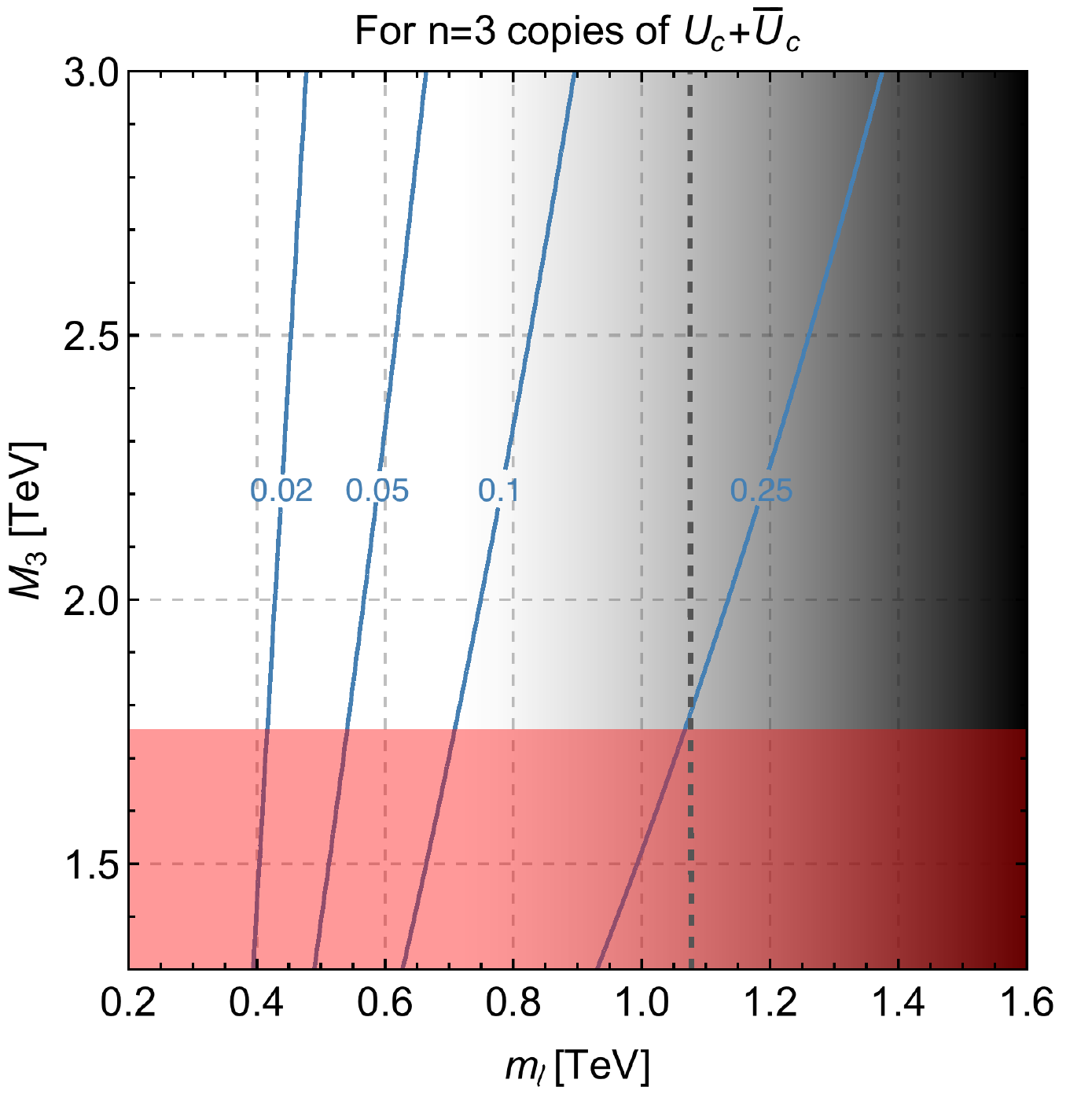}\quad\includegraphics[scale=0.5205]{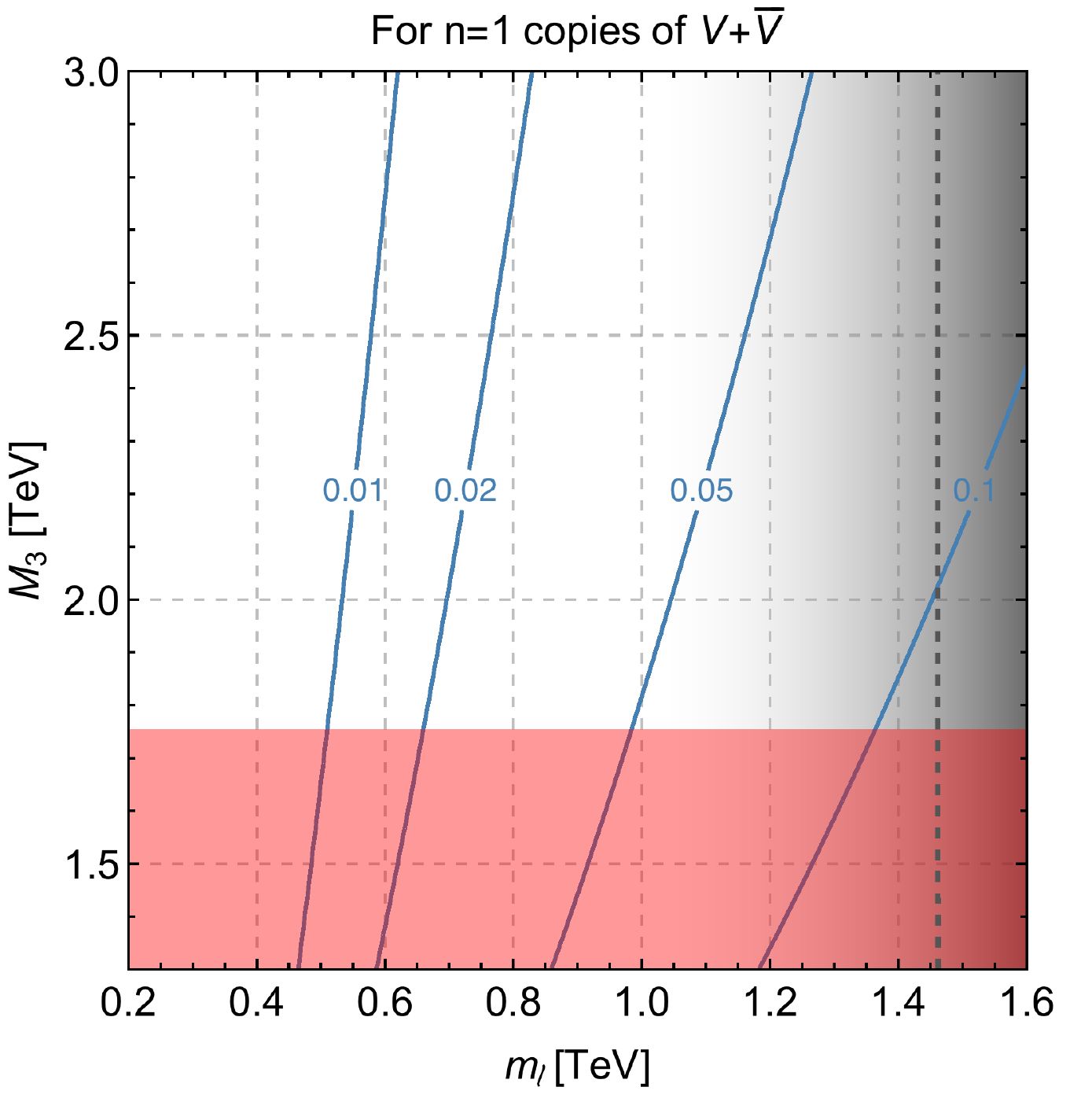}  \end{center}
\caption{ 
\emph{Contour lines of constant $\lmin$ (blue) fitting the signal, for $\sigma_{\gamma\gamma} = 8\fb$, in the plane of gluino and light messenger masses.  We have shaded in  red the gluino mass exclusion and in a gray gradient the non-perturbativity region $\geff \gtrsim \geff^*$. The dashed curve corresponds to the isocurve $\geff=\geff^*$.} }
\label{example2}
\end{figure}

To stick to known SM representations, let us consider as a first example $n$ copies of $U^c+\overline{U}^c$, where $U^c$ has the same quantum numbers as the SM up quark singlets. The Dynkin indices are $(N_1,N_2,N_3) = n\, (8/5,0,1)$. As the perturbativity of $g_1$ forces $n\leq 3$, we can take $n=3$. We then have $N_3 = 3$ in \eq{N3NC}, which, for $M_3-\Delta M_3 = m_l$, is compatible with $\geff < \geff^*$ when $\lmin \lesssim 0.4$. As for \eq{NgammaNC}, the three copies of $U^c+\overline{U}^c$ give $N_\gamma = 8$, which is compatible with $\geff \lesssim \geff^*$.

A second example, involving a SM representation included in the adjoint of SU(5), is the case of $n$ copies of $V+\overline{V}$, where $V$ is a fundamental of SU(3)$_c$ and SU(2)$_L$ with hypercharge $Y = - 5/6$. The Dynkin indices are $(N_1,N_2,N_3) = n\, (5,3,2)$. As the perturbativity of $g_1$ forces $n\leq 1$, we take $n=1$. We then have $N_3 = 2$ in \eq{N3NC}, which, for $M_3-\Delta M_3 = m_l$, is compatible with $\geff < \geff^*$ when $\lmin \lesssim 0.25$. And in \eq{NgammaNC} we have $N_\gamma = 34/3\approx 11$, which is compatible with $\geff \lesssim \geff^*$. 

In Fig.~\ref{example2} we show two plots  with contour lines in the $(M_3,m_l)$-plane corresponding to different values of $\lmin$ fitting the signal. The plots are  done with the one-loop formulas given in the Appendix. On the left hand side for $n=3$ copies of $U^c+\bar U^c$ fields, and as discussed above for this choice of parameters the model is on the edge of non-perturbativity, shown with a  dashed curve. On the right hand side the plot is done for one set of $V+\bar V$ messengers  in the adjoint of SU(5). 

The two above examples have different predictions for the $pp\to \gamma Z, ZZ, WW$ rates. The first example ($U^c+\bar{U}^c$) predicts them to be well below the present limits: $\Gamma_{ZZ}/\Gamma_{\gamma\gamma} \approx 0.08$, $\Gamma_{Z\gamma}/\Gamma_{\gamma\gamma} \approx 0.6$. The second example ($V+\bar V$) predicts higher rates, but also below the present experimental limits, with an accidental suppression of the $Z\gamma$ rate: $\Gamma_{ZZ}/\Gamma_{\gamma\gamma} \approx 1.3$, $\Gamma_{Z\gamma}/\Gamma_{\gamma\gamma} \approx 0.02$, $\Gamma_{WW}/\Gamma_{\gamma\gamma} \approx 2.8$. The previous ratios have been obtained using the formulas in the Appendix. 

Neither of the previous examples preserves the successful gauge coupling unification achieved in the MSSM. The simplest way to preserve unification is to add extra fields (non near-critical) so that the messengers form complete SU(5) multiplets. This is possible only for the second example. The first example would require completing the $3\times(U^c+\overline{U}^c)$ into $3\times ({\bf 10}+{\bf \overline{10}})$ of SU(5). In this case, however, the total Dynkin would be $N_1 = N_2 = N_3 = 9$, well above what perturbative gauge unification requires.

The second example, on the other hand, requires completing $V+\overline{V}$ to a full SU(5) adjoint, with $N_1 = N_2 = N_3 = 5$, at the boundary of gauge coupling unification. In this case, the fields $V+\overline{V} %\sim (\bf{3},\bf{2})_{\pm 5/6}
$ are accompanied by an adjoint of SU(3)$_c$, $\Sigma$, an adjoint of SU(2)$_L$, $W$, and a singlet. If $X$ is a singlet of SU(5), the $\lambda$ couplings of those fields are the same at the GUT scale, $\lambda_V = \lambda_\Sigma = \lambda_W$. The RGE running to low energies makes the triplet coupling $\lambda_W$ lower than $\lambda_V$, which prevents $V+\overline{V}$ from being near-critical (only the triplet can be). On the other hand, if $X$ is the singlet component of a SU(5) adjoint, then $\lambda_V  =   \lambda_W/6  = \lambda_\Sigma/4$ at the GUT scale. For perturbative values of the couplings at the GUT scale, $\lambda_V$ remains the smallest coupling during the whole running, and $V+\overline{V}$ can play the role of the near-critical fields.\footnote{The adjoint containing $V+\bar V$ also contains a gauge singlet. Its Yukawa coupling is likely to be the smallest one at low energy.  Therefore, when $V+\bar V$ is near critical, the singlet will be in the broken phase and develop a vev. This is not a problem, as the vev would not break the SM gauge symmetry.} In this scenario, the unification prediction for $\alpha_3(M_Z)$ is shifted by $\Delta \alpha_3 \approx -0.008$ with respect to the MSSM prediction, in the direction of the measured value.

Other solutions can be obtained in the context of flipped SU(5)~\cite{Barr:1981qv,Derendinger:1983aj} or using messenger spectra that preserve gauge coupling unification but are not in full SU(5) multiplets~\cite{Calibbi:2009cp}. 

In summary, it is possible to choose a messenger spectrum such that the bounds in eqs.~(\ref{eq:NgammaNC},\ref{eq:N3NC}) are satisfied with $\geff < \geff^*$ and that gauge couplings are perturbative up to the GUT scale, where they unify. The simplest example we found is the case of messengers forming a whole adjoint representation of SU(5), with the $Y\neq 0$ components near-critical and the $Y= 0$ components off near-criticality. 

\subsection{Strong coupling}

%So far, for the sake of predictivity, we have leaned towards a semi-perturbative regime for the light scalar messenger interactions. On the other hand, the sgoldstino interpretation of the diphoton anomaly in the near-critical regime definitely seems to prefer a non-perturbative regime for the scalar trilinear coupling, corresponding to $\geff = \lambda^2 M/m_l \sim 4\pi$. It is therefore worth speculating on the behaviour of the model in that regime. 

From the discussion of possible models in the previous Section, the following dichotomy emerges between UV and IR non-perturbativity. On one side, one can choose to have large representations or a large number of messenger fields. This allows to interpret the diphoton excess at the expense of Landau poles at some tens or hundreds of TeVs, thus having strong dynamics in the UV, but a weakly coupled EFT at TeV energies.

On the other side, one can avoid Landau poles by incorporating a smaller number of messenger fields, at the expense of strong dynamics in the IR, since the  trilinear  $\lmin$ has to be tuned close to criticality. 
%Such a non-perturbative regime corresponds to a large value of an irrelevant coupling ($\propto M$) and is therefore of quite a different nature than the non-perturbative regime one obtains for large values of marginal or irrelevant couplings. 
As an infrared effect, it does not give rise to Landau poles and it does not spoil the nice UV properties of supersymmetric theories.

In the IR non-perturbative regime, the numerical results showed above may receive large corrections. Moreover, the trilinear interaction leads to an attractive force between the light scalar messengers and, in  the IR strong coupling regime $\geff\gtrsim\geff^*$, one expects  a tower of bound states. In fact, a similar phenomenon is argued to happen in the MSSM if the trilinear interaction $A_t H \tilde q_L \tilde u_R$ becomes strong~\cite{Giudice:1998dj}. Thus we expect that the light scalar messengers form an S-wave color-singlet bound state $S_b$. The resonance $S_b$ would be a tightly bound state, as the binding energy is controlled by $m_l$. And it would be a true bound state, as the formation time is controlled by the inverse of the binding energy, and the decay by perturbative QCD interactions. Then, the bound state $S_b$ would mix with the sgoldstino and, since the  constituents of $S_b$ are colored, this would give rise to a direct coupling between gluons and the physical state.
% $\sim\cos \alpha |s\rangle + \sin\alpha |S_b\rangle$.
To our knowledge, the details of the phenomenology and the interplay of the sgoldstino and the possible singlet resonance is far from settled. It would be interesting to further explore such phenomenology, perhaps through lattice techniques, especially if in the future the sgoldstino scenario near criticality will gain further support from the experiments.

\section{Speculations on the origin of the near-critical regime}
\label{sec:speculations}

\begin{figure}[t]
\begin{center}
  \includegraphics[scale=0.60]{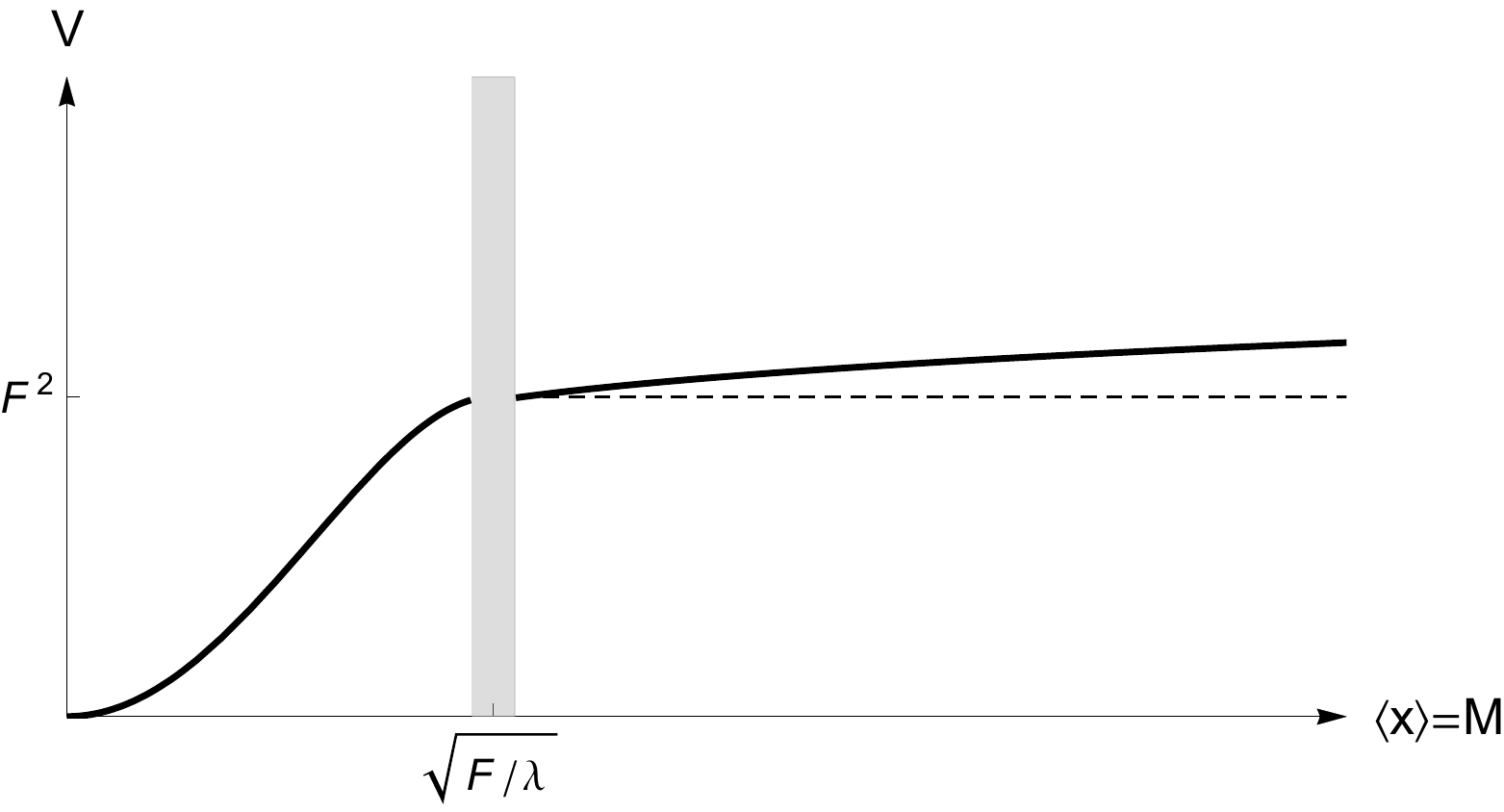}  \end{center}
\caption{ 
\emph{Schematic representation of the one-loop corrected potential in Sect.~\ref{sec:speculations} as a function of $\langle x \rangle$. The tree-level potential in the pseudo-flat direction is plotted with a dashed line. The shaded region corresponds to the region of near-criticality (the size is largely exagerated). }}
\label{example3}
\end{figure}

The near-critical regime requires an apparent fine-tuning making $m_l \ll \lambda M$ and enhancing $\geff$ in \eq{geff}. In this Section, we speculate on a possible connection between a dynamical origin of near-criticality and the strong regime it induces. 

Let us consider the simplest possible completion of the basic model building block considered in the previous Section, \eq{W}. The gauge group is $G_\text{SM}$. Besides the MSSM ones, the chiral superfields are $X$ and a vectorlike set of messengers $\Phi+\bar{\Phi}$. For simplicity, we take $\Phi$ in a single irreducible SM representation. The superpotential is
\begin{equation}
W = W_\text{MSSM} + \lambda X \Phi \bar\Phi - F X ,
\label{eq:Wspe}
\end{equation}
where $\lambda >0$, $F >0$ with no loss of generality. While the above system is simple and well known, we are not aware of a thorough discussion of its behaviour near the critical point. 

Let us remind the main features of the above system. We expand $X$ as in \eq{X}. Then the system has a supersymmetric minimum for $\vev{x} = 0$, $|\vev{\Phi}| = |\vev{\bar\Phi}| = \sqrt{F/\lambda}$. The system has two phases, controlled by the scalar vev of $x$, $\vev{x} = M$, which can also be taken real and non-negative. When $M > \sqrt{F/\lambda}$, the critical point, the messengers have no vev, the gauge symmetry is unbroken, and supersymmetry is broken by the $F$-term vev of $X$, $\vev{f} = F$. In such an unbroken phase, the tree level potential is flat, $V = F^2$. On the other hand, when $M < \sqrt{F/\lambda}$,  both messengers get a vev of size $|\vev{\Phi}| = |\vev{\bar\Phi}| = \sqrt{F/\lambda - M^2}$ and the potential, as a function of $M$, is 
\begin{equation}
V = F^2-(F - \lambda M^2)^2 ,
\label{eq:VM}
\end{equation}
which forces $M = 0$, where the potential has its absolute minimum, the gauge symmetry if broken, and supersymmetry is unbroken. 

On the left of the critical point, $M < \sqrt{F/\lambda}$, the tree level potential provides a sufficiently accurate description (except perhaps very near the critical point, where it is nearly flat). On the right side, on the other hand, the flat direction for $M >  \sqrt{F/\lambda}$ is lifted by the one-loop correction to the potential, which acquires a positive slope and let $M$ slide towards the critical point, see Fig.~\ref{example3}.

In a region around the critical point, though, the system hides a non-perturbative regime, triggered by the growth of the coupling in \eq{geff}. Interestingly, this happens only in a tiny region, characterised by 
\begin{equation}
\lambda M^2 - F \lesssim \frac{F}{(4\pi)^2} . 
\label{eq:window}
\end{equation}
We can then speculate on the possibility that the effective potential generated by strong interactions has a negative slope in some point of the near-critical region. If that were the case, a metastable minimum would be generated for the potential in the near-critical region, thus providing a dynamical origin for the apparently fine-tuned closeness of $\lambda M^2$ and $F$ (and for the origin of supersymmetry breaking). But this is of course just wishful thinking. Still, we consider worth pointing out the existence of an ephemeral, but potentially interesting, non-perturbative regime in a simple and well known model, which might deserve further investigation. 

If it stabilised in the in the near-critical region, the system we considered has an obvious, spontaneously broken $R$-symmetry. The phase of $x$ would play the role of the $R$-axion and would be massless, up to supergravity $R$-breaking corrections. In the next Section, we will make a few considerations on the possible implications of such a light $R$-axion, in a more general context.

\section{The $R$-axion}
\label{sec:collimated}

The discussion in Sect.~\ref{sec:origin} assumed that the only relevant interactions of the diphoton resonance are the ones with the messengers in \eq{W}. On the other hand, the hidden sector dynamics can in principle give rise to alternative decay channels that could compete with the photon and gluon ones, thus affecting our phenomenological analysis. However, it is not unreasonable to assume that most of the hidden degrees of freedom are irrelevant because characterised by large, $\mathcal{O}(\sqrt{F})$ masses. 

Even if that is the case, there are two possible decay channels that can not be ignored. 
First, the SUSY breaking mass of the sgoldstino $m_s^2/F^2 |X|^4\big|_{D}$ leads to a decay of the sgoldstino into two goldstinos. Such  a decay is negligible with respect to $\Gamma(s\rightarrow\gamma\gamma)$, see eq.~(\ref{partial01}), in the regions of parameter space that we consider. Then, a wide class of supersymmetry breaking models predicts the existence of a light degree of freedom, the $R$-axion, which could be relevant. In this Section, we shortly outline the possible role of a light $R$-axion in the diphoton phenomenology. 
The $R$-symmetry plays a central role in most supersymmetry breaking models. If present, its spontaneous breaking is welcome to allow for Majorana gaugino masses. The $R$-axion is the Goldstone associated to such a spontaneous breaking. As such, it is massless, up to the small explicit breaking of the $R$-symmetry provided by supergravity~\cite{Bagger:1994hh}. Which give it a mass that, in our parameter space, is $\ord{100\MeV}$~\cite{Goh:2008xz}, with a non-negligible dependence on the hidden sector dynamics. If $X$ is the only source of spontaneous $R$-symmetry breaking, as in the toy model in the previous Section, the $R$-axion is the phase $a$ of its scalar component, $x = (M + s/\sqrt{2})e^{i a/(\sqrt{2}M)}$ (where with an abuse of notations we are using the same notations for the fields $s$, $a$ that we used for their linearised versions in \eq{X}). In general, the $R$-axion will have at least a component in $a$, if $X$ has a non-vanishing $R$-charge. For simplicity, and to maximise the role of the $R$-axion in the diphoton phenomenology, we will assume that the $R$-axion coincides with $a$. 

The radial component $s$, i.e.\ the $\sim 750\GeV$ resonance, can then decay into two $R$-axions, with a partial width
\begin{equation}
\Gamma(s\to aa) = \frac{1}{64 \pi} \frac{m^3_s}{M^2} .
\label{eq:Rwidth}
\end{equation}
Depending on the subsequent fate of the $R$-axion and on the relative size of the above width and the $gg$, $\gamma\gamma$ ones, the above channel, if present, can affect the discussion in Sect.~\ref{sec:origin}. 

The mass of the $R$-axion is in the ballpark of the pion mass. If $m_a > 2m_\pi$, it will dominantly decay into two pions.~\footnote{The $R$-axion decay into two gravitinos is very much suppressed in our region of parameter space~\cite{Bagger:1994hh,Goh:2008xz}.}
If $m_a < 2m_\pi$,  the relevant channels  are $R$-axion decays  into two photons, electrons or muons. 
The lepton decays are proportional to $m_a (m_f/M)^2$, see Ref.~\cite{Goh:2008xz}, and can be suppressed with respect to the decay into photons for the messenger mass scales that we  consider. Then, the decay into two photons is in principle relevant, as the values of the $R$-axion mass just happens to be in the window in which the two photons are collimated enough to be seen at the LHC as a single photon~\cite{Aparicio:2016iwr,Dobrescu:2000jt,Knapen:2015dap,Agrawal:2015dbf,Chang:2015sdy,Bi:2015lcf,Chala:2015cev,Domingo:2016unq,Ellwanger:2016qax,Chiang:2016eav,Bi:2016gca,Tsai:2016lfg}. As a consequence, the decay $s\to aa$ could in principle also account for the diphoton signal, through the subsequent decay of the $R$-axions into two collimated photons. Unfortunately, the lifetime of the $R$-axion is too long for the decay to take place before hitting the detector. In order for the two photons to be collimated enough, the mass of the $R$-axion should conservatively be below $200\MeV$, and this is already in tension the possibility that the decay is induced by dynamics at the TeV scale~\cite{Aparicio:2016iwr}. As in our case the decay is induced by dynamics at the few $\ord{10\TeV}$ scale (unlike the resonance $s$, the $R$-axion has no trilinear couplding to the light scalar messengers, \eq{trilinear}), there is no chance that it will be fast enough to give rise to the diphoton signal. Except if the $R$-axion mass happens to be very close to the pion mass (or the $\eta$ mass, but that value of the masses might be too large~\cite{Aparicio:2016iwr})~\cite{Domingo:2016unq,Ellwanger:2016qax}. In the latter case, a non-negligible mixing with the pion would allow the $R$-axion to decay as a pion, well before hitting the detector. In summary, the fate of the $R$-axion is either to contribute to the invisible width of $s$ or, in a fine-tuned window for its mass, to contribute to the diphoton signal. 

As for the relative size of the widths, the $gg$ and $\gamma\gamma$ widths are suppressed by a loop factor, compared to \eq{Rwidth}, but the $aa$ width is suppressed by a higher scale, $M^2$ versus $m_l^2$. The relative sizes of the widths then depends on the specific values of the parameters one considers, and is controlled by $\lambda M /m_l = \geff/\lambda$,
\begin{equation}
\frac{\Gamma(s\to gg)}{\Gamma(s\to aa)} \approx 
\frac{\alpha^2_3}{36\pi^2} \fracwithdelims{(}{)}{\geff}{\lambda}^4 \bar N_3^2,
\qquad
\frac{\Gamma(s\to \gamma\gamma)}{\Gamma(s\to aa)} \approx 
\frac{\alpha^2}{288\pi^2} \fracwithdelims{(}{)}{\geff}{\lambda}^4 \bar N_\gamma^2. 
\label{eq:relative}
\end{equation}
Assuming for simplicity that the $R$-axion mass is below $200\GeV$, we have three regimes (assuming $N_\gamma \lesssim 34\, N_3$). 

\begin{itemize}
\item $80/\sqrt{N_\gamma}\lesssim (\geff/\lambda)$ \\[2mm]
In this regime, the decay in $R$-axions is subdominant to both the decay into $gg$ and $\gamma\gamma$. Therefore, it does not affect the discussion in Sect.~\ref{sec:origin}. 
\item $14/\sqrt{N_3} \lesssim (\geff/\lambda) \lesssim 80/\sqrt{N_\gamma}$  \\[2mm]
In this regime, the decay width in $R$-axions is larger than the decay width in $\gamma\gamma$, but not of the decay width in $gg$. Therefore, it does not affect the discussion in Sect.~\ref{sec:origin}, except in the fine-tuned window in which it mixes with the pion. In such a case, it gives the dominant contribution to the diphoton signal, and the $\saa$ rate determines $\Gamma(s\to aa)$,
\begin{equation}
\Gamma(s\to aa) \approx 0.4\MeV \fracwithdelims{(}{)}{\saa}{8\fb} 
\quad\Rightarrow \quad
M \approx 69\TeV \fracwithdelims{(}{)}{8\fb}{\saa}^{1/2} .
\label{eq:GminR2}
\end{equation}
\item $(\geff/\lambda) \lesssim 14/\sqrt{N_3}$  \\[2mm]
In this regime, the decay width in $R$-axions is larger than both the decay widths in $gg$ and $\gamma\gamma$. The diphoton signal is then suppressed, compared to what found in Sect.~\ref{sec:origin}, which should be avoided. Except in the fine-tuned region in which the $R$-axion decay into two photons is enhanced by the mixing with the pion, in which case the $\saa$ rate determines $\Gamma(s\to gg)$,
\begin{equation}
\Gamma(s\to gg) \approx 0.4\MeV \fracwithdelims{(}{)}{\saa}{8\fb} 
\quad\Rightarrow \quad
\frac{\geff}{\geff^*} \approx \frac{0.2}{\bar N_{3}} \fracwithdelims{(}{)}{m_l}{\text{TeV}} \fracwithdelims{(}{)}{\sigma_{\gamma\gamma}}{8\fb}^{1/2} . 
\label{eq:GminR3}
\end{equation}
\end{itemize}

\section{Sfermion masses and $D$-terms}
\label{sec:D}

In the previous Section, we have ignored the MSSM sfermions. On the other hand, if the only contribution to their mass was the minimal gauge mediation two-loop contribution that follows from \eq{W}, we would expect the colored sfermions to be lighter than the gluino, in which case they would play a role at least in forcing the whole spectrum to be heavier in order to pass the experimental bounds. In this Section we show that i) it is indeed easy to split the spectrum and make the sfermions parametrically heavier than the gauginos, so that they do not play a role in the diphoton phenomenology, and ii) the model building ingredients needed to make them heavy modify the dynamics discussed in Sect.~\ref{sec:origin}, but have a minor impact on the conclusions. 

In order to make the sfermions parametrically heavier than the gauginos it suffices to make both $X$ and the sfermions charged under a (non-anomalous) U(1)$_X$ gauge factor. As they couple to $X$, the messengers are then also charged under $X$. The vev of $X$ breaks U(1)$_X$ and supersymmetry at the same time. As a consequence, the sfermions (and the scalar messenger, or ``smessengers'') get tree-level soft masses from the U(1)$_X$ $D$-term, parametrically larger than the loop-induced gaugino masses. In the near critical regime, in which, as we will see, we still have $F\sim \lambda^2 M$, the $D$-term contribution to the soft mass of the scalar $\varphi$ (sfermion or smessenger), $m^2_\varphi$, is of the order of the heavy messenger scale, 
\begin{equation}
m^2_\varphi = q_\varphi\, g_X D \sim \frac{F^2}{M^2} \sim \lambda^2 M^2 = M^2_m,
\label{eq:mmD}
\end{equation}
where $q_\varphi$ is the U(1)$_X$ charge of the scalar field, $g_X$ the gauge coupling, and $D$ the $D$-term. With $M_m$ in the (10--100)$\TeV$ range, we are dealing with a simple realisation of the split supersymmetry spectrum~\cite{ArkaniHamed:2004fb,Giudice:2004tc,ArkaniHamed:2004yi}~\footnote{A family-dependent assignment of U(1)$_X$ charges would give rise to a simple realisation of natural susy spectrum.}. 

The large soft terms have a relevant impact on both sfermions and smessengers dynamics. The sfermions are too heavy to affect the diphoton phenomenology, as desired. In order for them not to be tachyonic, their U(1)$_X$ charges need to have the same sign as the $D$-term, say positive for definitess. As U(1)$_X$ is non-anomalous, the supertrace must vanish, and the tree-level scalar soft masses must add up to zero. The positiveness of the MSSM sfermion masses hence forces negative soft mass terms for some scalars. This had long been considered as an obstacle to tree-level mediation of supersymmetry breaking in non-anomalous, renormalizable theories. But it is not: the messengers are anyway forced to have (overall) negative soft mass terms, as they couple to a positively charged field (see below). That does not make them tachyonic, as the negative soft mass term is compensated by the positive, supersymmetric, mass term. And their soft mass can compensate the positive sfermion soft masses. Such a class of models, in which the sfermions with negative soft masses needed to satisfy the supertrace constraint get a large, positive, supersymmetric mass term from U(1)$_X$ breaking and play the role of chiral messengers of minimal gauge mediation has been studied in Refs.~\cite{Nardecchia:2009ew,Nardecchia:2009nh,Caracciolo:2012de}.
% see also Refs.~\cite{Monaco:2011fe,Monaco:2013poa}. 
The compensation, i.e.\ anomaly cancellation, can arise automatically if the U(1)$_X$ is embedded in non-abelian gauge groups. 

Let us now consider the impact of the new $D$-term contributions on the smessenger dynamics, and show that the conclusions obtained in Sect.~\ref{sec:origin} are unchanged. Let $q_X = 1$ be the charge of $X$\footnote{Up to normalisation, $q_X = \pm 1$. If $X$ is the dominant source of supersymmetry breaking, $D>0$ is obtained for $q_X =1$.}, $-q$, $-\bar q$ the charges of $\Phi$, $\overline{\Phi}$ (neglecting again the messenger flavour index $i$), with $q+\bar q = 1$, so that the total messenger soft mass, $m^2_\phi + m^2_{\bar\phi} = -g_X D$, is   negative. The messengers are then chiral under U(1)$_X$, which ``protects'' their masses, in the same sense in which the electroweak symmetry ``protects'' the SM fermion masses. \Eq{mmass} becomes
\begin{equation}
-\mathcal{L}_\text{mess}^{(2)} = 
\left(
\lambda M \psi \bar \psi + \text{h.c.}
\right) +
(\phi^\dagger, \bar\phi )
\begin{pmatrix}
\lambda^2 M^2 -q g_X D & \lambda F \\
\lambda F & \lambda^2 M^2 -\bar q g_X D
\end{pmatrix}
\begin{pmatrix}
\phi \\
\bar{\phi}^\dagger
\end{pmatrix} ,
\label{eq:mmassD}
\end{equation}
where all terms in the smessenger mass matrix are of the same order. In order to avoid tachyons, we need $\lambda^2M^2 \geq q g_X D$ (by assumption larger than $\bar q\, g_X D$) and 
\begin{equation}
\lambda^2 F^2 \leq M^2_\phi M^2_{\bar\phi} ,
\label{eq:tachyonsD}
\end{equation}
where
\begin{equation}
\begin{aligned}
M^2_{\phi} &= \lambda^2 M^2 - q g_X D , \\
M^2_{\bar \phi} &= \lambda^2 M^2 - \bar q g_X D .
\end{aligned} 
\label{eq:MMphi}
\end{equation}
Near criticality (and a small smessenger mass $m^2_l \ll M^2_m$) is obtained when the condition in \eq{tachyonsD} is just satisfied, with $F^2$ just below the upper limit. Note that the near-critical regime cannot be associated to a fine-tuned cancellation in $M^2_{\phi}$, as that would imply $F\ll \lambda M^2$ and $D \sim F^2/M^2 \ll \lambda^2M^2$. As a consequence, $F\sim \lambda M^2$. The heavy and light mass eigenstates $\phi_h$, $\phi_l$ have now mass
\begin{equation}
m^2_{h,l} = \frac{M^2_\phi + M^2_{\bar\phi}}{2} \pm 
\left[
\fracwithdelims{(}{)}{M^2_\phi - M^2_{\bar\phi}}{2}^2 + \lambda^2 F^2
\right]^{1/2}
\label{eq:masshlD}
\end{equation}
and are given by 
\begin{equation}
% \left\{
\begin{aligned}
\phi &= \cos\theta \, \phi_h - \sin\theta \, \phi_l \\
\bar{\phi}^* &= \sin\theta \, \phi_h + \cos\theta \, \phi_l
\end{aligned}
% \right. 
,
\qquad 
\sin 2\theta = \frac{2 \lambda F}{m^2_h - m^2_l} .
\label{eq:hl}
\end{equation}

The diphoton signal is not affected by the $D$-term contributions to the smessenger masses. The angle describing the mixing in the smessenger sector, now possibly different from $\pi/4$, does not enter the relevant trilinear interactions, which still have the form in \eq{trilinear}. The decay widths are therefore unchanged (for given $m_l$), in the near critical limit in which the light smessenger exchange dominates the diphoton signal. In particular, the effective coupling of the resonance to the light smessenger is still given by \eq{geff}. 

On the other hand, the $D$-term has a mild effect on the relation of the gluino mass to the smessenger masses. We have in fact
\begin{equation}
\begin{gathered}
M_a = \frac{\alpha_a}{4\pi} \frac{F}{M} N_a \, g \left(
\frac{m^2_l}{M^2_m}, \frac{m^2_h}{M^2_m}
\right) , \\[2mm]
g(x_1,x_2) = \frac{2}{x_1-x_2} 
\left(
\frac{x_1 \log x_1}{x_1-1} - \frac{x_2 \log x_2}{x_2-1}
\right) ,
\label{eq:gauD}
\end{gathered}
\end{equation}
and, in the near-critical regime, 
\begin{equation}
M_a = \frac{\alpha_a}{4\pi} M_m N_a \frac{\sqrt{(1-qr)(1-\bar q r)}}{1-r} 2\log(2-r) ,
\quad\text{with}\quad
r = \frac{g_X D}{M^2_m} ,
\label{eq:gauNCD}
\end{equation}
and $M_m = (\geff/\lambda) m_l$, as before. For $r\to 0$ (and $\Delta M_3 =0$), \eq{gauNCD} reproduces \eq{M3NC}. Numerically, for given $m_l$ and $\geff$, the presence of $D$-terms, i.e.\ of a non-zero $r$, gives only slightly lower values of gluino masses. We therefore conclude that in the discussion of Sect.~\ref{sec:origin} the sfermions can be easily made heavy, without significantly modifying the conclusions about the possibility to fit the signal within the constraints. On the other hand, the mixing in \eq{hl}, if not maximal, could induce a decay $s\to hh$. In order to avoid that, it suffices to give $\Phi$ and $\bar\Phi$ the same charge under U(1)$_X$, $q = \bar q = 1/2$. This is certainly the case in the $V+\bar V$ example of Sect.~\ref{sec:examples}, as both fields originate from the same adjoint.

\section{Summary and outlook}
\label{sec:conclusions}

We have revisited the possibility to associate the recently reported diphoton excess to the production of a sgoldstino of about $750\GeV$. In this context, the new degree of freedom is not an ad hoc degree of freedom, it is ordered by the need to break supersymmetry, in the context of a theory with its own appeal; and the experiment measures the scale of mediation of supersymmetry breaking, which turns out to be very low, $\ord{100\TeV}$ or less. 

We assumed that supersymmetry breaking, and thus the sgoldstino resonance, is coupled to the MSSM fields through gauge mediation, which is appropriate at such low scales. The messenger superfields then provide the additional degrees of freedom needed for the decay and production of the resonance. 

We showed that the experimental bounds on gaugino masses force the messenger scale $M_m$ to be $\sim (10$--$100)\TeV$ and thus make the sgoldstino contribution to the diphoton excess unobservable, for a reasonable messenger content; except in a small region of the parameter space near the critical point beyond which the messengers get a vev, $F\approx \lambda M^2$. 

The phenomenology in this thin, nearly-critical region drastically departs from the standard gauge mediation picture. One (or more) of the scalar messengers becomes much lighter than the heavy messenger scale. It can therefore lie at the TeV scale, as needed to account for the diphoton excess. At the same time, when the messenger becomes much lighter than $M_m$, its effective trilinear coupling gets enhanced by a factor $\lambda M_m/m_l$, where $m_l$ is the light messenger mass, thus further helping to account for the excess. When the enhancement becomes very large, the system enters a strongly interacting regime. It is then not possible to further raise the gain through a larger hierarchy between the heavy and light messengers. 

The IR non-perturbativity found at small $m_l\ll M_m$, associated to a large irrelevant coupling, is of quite a different nature than the usual UV non-perturbativity associated to irrelevant or marginal coupling. As an infrared effect, it does not give rise to Landau poles and it does not spoil the UV properties of supersymmetric theories. A quantitative analysis of the possibility to account for the diphoton effect showed a dychotomy between those two   regimes. On the one hand, it is possible to account for the diphoton excess while avoiding the IR strong coupling by using a large enough set of messenger fields. This however forces Landau poles well below the GUT scale, and thus strong dynamics in the UV. On the other hand, it is possible to maintain the theory perturbative in the UV (up to the GUT scale) by having a lower number of messengers, but that forces a large trilinear coupling and induces strong dynamics in the IR. 

The IR regime is particularly intriguing. It requires the system to be in a fine-tuned near-critical region, where the determinant of the scalar messenger mass matrix is small. In the context of the simplest possible structure of the hidden sector, $W_\text{hidden} = F X$, we observed that the near critical region is located on one end of the metastable flat direction associated to $X$, i.e.\ around the critical point, before the cascade to the supersymmetric minimum. The shape of the loop corrected effective potential along the flat direction is well known, it slowly pushes $X$ towards the cascade. On the other hand, because of the non-perturbative regime arising there, the shape of the potential in the near critical region is not obvious. The obvious speculation is then that a metastable minimum could form in the near-critical region, thus providing a dynamical origin for the apparent fine-tuning we need, and for the origin of supersymmetry breaking. But this is of course just wishful thinking. In any case, an investigation of that region with non-perturbative methods would be welcome. 

Back to phenomenology, we did not aim at accounting for a possibly large width of the resonance, relying of the presently uncertain experimental situation. In particular, the known interpretation of an apparent width in terms of the production of two resonances close in mass, identified with the scalar and pseudoscalar components of the sgoldstino, is not available here, as the pseudoscalar component has no (enhanced) trilinear coupling to the messengers. 

In passing, we have commented on the role of a possible $R$-axion in the analysis of the diphoton excess. In this setup, the $R$-axion mass is in the ballpark of the pion mass. That is the window in which the decay of the sgoldstino into two $R$-axions, followed by the subsequent decay of each $R$-axion into two collimated photons, would mimic a diphoton signal. On the other hand, the lifetime of the $R$-axion would be too long for it to decay before the detector, except when a sizeable mixing with the pion arises. 

Finally, we have shown that it is possible to give the sfermions a mass parametrically larger than the gauginos ones, so that they have no impact on our discussion, without altering our conclusions. 

\section*{Acknowledgments}

We thank Aleksandr Azatov, Jos\'e Ram\'on Espinosa, Edward Hardy, and Giovanni Villadoro for useful discussions. The work of A.R.\ was supported by the ERC Advanced Grant no. 267985 ``DaMESyFla''.

\appendix

\section{Partial widths}

In this Appendix we give the one loop expression for the partial decay widths of $s$ and $a$ into $gg$, $\gamma\gamma$, $ZZ$, $Z\gamma$, $WW$, neglecting the mass of the massive gauge bosons. 

\bea
\Gamma(s\to gg) &=& m_s \frac{8 \alpha^2_3}{(4\pi)^3}
\left|
\sum_i \frac{\lambda_i}{\sqrt{2}} N_{3,i} \sqrt{x_i} 
\left[
P(x_i) + \frac{F(x_{i,l}) + F(x_{i,h})-2F(x_i)}{2}
\right]
\right|^2
\\[.2cm]
\Gamma(a\to gg) &=& m_a \frac{8 \alpha^2_3}{(4\pi)^3}
\left|
\sum_i \frac{\lambda_i}{\sqrt{2}} N_{3,i} \sqrt{x_i} 
\, P(x_i) 
\right|^2 
\eea
\bea
\Gamma(s\to \gamma\gamma) &=& m_s \frac{\alpha^2}{(4\pi)^3}
\left|
\sum_i \frac{\lambda_i}{\sqrt{2}} N_{\gamma,i} \sqrt{x_i} 
\left[
P(x_i) + \frac{F(x_{i,l}) + F(x_{i,h})-2F(x_i)}{2}
\right]
\right|^2
\\[.2cm]
\Gamma(a\to \gamma\gamma) &=& m_a \frac{\alpha^2}{(4\pi)^3}
\left|
\sum_i \frac{\lambda_i}{\sqrt{2}} N_{\gamma,i} \sqrt{x_i} 
\,P(x_i) 
\right|^2
\eea
\bea
\Gamma(s\to ZZ) &=& m_s \frac{\alpha^2}{(4\pi)^3}
\left|
\sum_i \frac{\lambda_i}{\sqrt{2}} N_{Z,i} \sqrt{x_i} 
\left[
P(x_i) + \frac{F(x_{i,l}) + F(x_{i,h})-2F(x_i)}{2}
\right]
\right|^2
\\[.2cm]
\Gamma(a\to ZZ) &=& m_a \frac{\alpha^2}{(4\pi)^3}
\left|
\sum_i \frac{\lambda_i}{\sqrt{2}} N_{Z,i} \sqrt{x_i} 
\,P(x_i) 
\right|^2
\eea
\bea
\Gamma(s\to Z\gamma) &=& m_s \frac{2\alpha^2}{(4\pi)^3}
\left|
\sum_i \frac{\lambda_i}{\sqrt{2}} N_{Z\gamma,i} \sqrt{x_i} 
\left[
P(x_i) + \frac{F(x_{i,l}) + F(x_{i,h})-2F(x_i)}{2}
\right]
\right|^2
\\[.2cm]
\Gamma(a\to Z\gamma) &=& m_a \frac{2\alpha^2}{(4\pi)^3}
\left|
\sum_i \frac{\lambda_i}{\sqrt{2}} N_{Z\gamma,i} \sqrt{x_i} 
\,P(x_i) 
\right|^2
\eea
\bea
\Gamma(s\to WW) &=& m_s \frac{2\alpha^2_2}{(4\pi)^3}
\left|
\sum_i \frac{\lambda_i}{\sqrt{2}} N_{W,i} \sqrt{x_i} 
\left[
P(x_i) + \frac{F(x_{i,l}) + F(x_{i,h})-2F(x_i)}{2}
\right]
\right|^2
\\[.2cm]
\Gamma(a\to WW) &=& m_a \frac{2\alpha^2_2}{(4\pi)^3}
\left|
\sum_i \frac{\lambda_i}{\sqrt{2}} N_{W,i} \sqrt{x_i} 
\,P(x_i) 
\right|^2.
\eea
In the above expressions, $F$ is the (off-shell) scalar loop function of $s$, $P$ is the (off-shell) fermion loop function of $a$, and the fermion loop function of $s$, $S$, has been expressed in terms of the previous two,
\begin{equation}
P(x) = \arctan^2\frac{1}{\sqrt{x-1}},
\qquad
F(x) = x P(x) -1 ,
\qquad
S(x) = P(x) - F(x). 
\label{eq:loopfunctions}
\end{equation}
The arguments of the loop functions are 
\begin{equation}
x_i  = 4\, \frac{\lambda^2_i M^2}{m^2_{s,a}},
\qquad
x_{ih,il} = 4\, \frac{m^2_{ih,il}}{m^2_{s,a}} . 
\label{eq:arguments}
\end{equation}
Finally, the Dynkin coefficients are
\begin{equation}
\begin{aligned}
N_\gamma &= \frac{5}{3} N_1 + N_2, \\
N_Z &= \frac{5}{3} \tan^2\theta_W N_1+ \cot^2 \theta_W N_2, \\
N_{Z\gamma} &= \frac{5}{3} \tan\theta_W N_1- \cot \theta_W N_2, \\[2mm]
N_W &= N_2 ,
\end{aligned}
\label{eq:Dcoeff}
\end{equation}
in terms of the SM Dynkin indices $N_{1,2,3}$.

% ====================================== START Bibliography
%\nocite{*}
\bibliographystyle{JHEP}
\bibliography{refs}
% ====================================== END Bibliography

\end{document}